\documentclass{PoS}
\usepackage{subcaption}

\vspace*{-7\baselineskip}
  \hbox{Belle Preprint 2017-04}
    \hbox{KEK Preprint 2016-65}
\title{Search for CP violation in $D$ decays to radiative and hadronic decays and search for rare $D$ decay at Belle}

\ShortTitle{Search for CP violation in $D^0$ decays to radiative and hadronic decays at Belle}

\author{\speaker{N. Dash}\thanks{On behalf of the Belle Collaboration}\\
        IIT BHUBANESWAR\\
        E-mail: \email{nd11@iitbbs.ac.in}}

\abstract{..........................\
          ...........................}

\FullConference{38th International Conference on High Energy Physics\\
		3-10 August 2016\\
		Chicago, USA}
\abstract{ 
Using the large data sample accumulated by the Belle experiment at KEKB asymmetry energy $e^+e^-$ Collider, we report the preliminary measurements of the branching fractions of the radiative decays $D^0 \rightarrow V\gamma$, where $V$ = $\phi,\bar{K^*}^0, \rho^0$. This is the first observation of the decay $D^0 \rightarrow \rho^0\gamma$ and also present the preliminary results of the first measurement of $CP$ asymmetry in these decays. We present the  preliminary result of the $CP$ asymmetry of the $D^0 \rightarrow K^0_S K^0_S$ decay, which is consistent  with no $CP$ violation and improves the uncertainty with respect to the previous measurement of this quantity by more than a factor of three. Finally, we present the results of the rare charm decay $D^0 \rightarrow \gamma\gamma$, resulting in the most restrictive upper limit to date.}

\begin{document}
\section{Introduction}
Charge-conjugation Parity Violation (CPV) in charmed meson decays has not yet been observed and is predicted in the Standard Model (SM) to be small (10$^{-3}$). An evidence for CPV in 2012 by LHCb ($-$0.82 $\pm$ 0.21 $\pm$ 0.11)\%~\cite{55} suggested a 3.5 standard deviation ($\sigma$), confirmed by CDF~\cite{66}, took people by surprise and revived the field. Measurements of $\Delta A_{CP}$ $(A _{CP}^{ D^{0} \rightarrow \pi^{+} \pi^{-}}$ $-$  $A_{CP}^{D^{0} \rightarrow K^{+} K^{-}}$) have been performed by LHCb, CDF, BaBar and Belle collaborations~\cite{33}-~\cite{lhcb21}. Recently LHCb updated  $\Delta A_{CP}$ result with a larger data sample and there is no evidence of non-zero asymmetry~\cite{lhcb21} . The combined result of $\Delta A_{CP}$ by HFAG~\cite{99} gives an agreement with no CP violation at 6.5\% CL. Though there is no current evidence of non-zero asymmetry, CPV in charm decays is investigated in other channels. It is interesting to measure CPV in $D^0$ decays to neutral final states such as : $V\gamma$ ($V$ = $\phi,\bar{K^*}^0, \rho^0$), $K^0_S K^0_S$. This report also presents a search for the rare charm decay $D^0 \rightarrow \gamma\gamma$, which has not been observed yet.
\paragraph*{} The $D^0$ candidates are selected as coming from the decay $D^{*+} \rightarrow D^0\pi^+_s$, where $\pi^+_s$ denotes the low-momentum "slow" pion. The charge of this slow pion reveals the flavor content of neutral $D$ meson (whether it is a $D^0$ or $\bar{D}^0$ ) at its production vertex ~\cite{61}. A stringent selection criterion is applied on the momentum of the $D^{*+}$ candidate in the $e^+e^-$ center-of-mass frame, $p^*(D^*)$, to suppress $D^{*+}$  coming from $B$ decays as well as to reduce the combinatorial background. The $D^{*+}$ mesons mostly originate from the $e^+e^- \rightarrow c\bar{c}$ process via hadronization, where the inclusive yield has a large uncertainty of 12.5\% ~\cite{pdg}. To avoid this uncertainty, we measure the branching fraction of signal decay mode with respect to the well measured mode as a normalization decay using the following relation:$\mathcal{B}_{sig} = \mathcal{B}_{norm} \times \frac{N_{sig}}{N_{norm}} \times \frac{\epsilon_{norm}}{\epsilon_{sig}}$, where $N$ is the extracted yield, $\epsilon$ the reconstructed efficiency and $\cal{B}$ the branching fraction for signal and normalization modes, respectively. For $\cal{B}$$_{norm}$, the world average value \cite{pdg} is used. Assuming the total decay width to be same for particles and antiparticles, the time-integrated $A_{CP} $ is given as: 

\begin{equation}
   A_{CP}=\frac{\Gamma (D^{0}\rightarrow f)-\Gamma (\bar D^{0}\rightarrow \bar f)}{\Gamma (D^{0}\rightarrow f)+\Gamma (\bar D^{0}\rightarrow {\bar {f}})}
\end{equation}

where, $\Gamma$ represents the partial decay width and $f$ is specific final state. The extracted raw asymmetry is given by:
 
\begin{equation}
  A_{\rm raw} = \frac{N(D^{0}\rightarrow f)- N(\bar D^{0}\rightarrow \bar f)}{N(D^{0}\rightarrow f)+ N(\bar D^{0}\rightarrow {\bar {f}})} =  A_{CP} + A_{FB} + A^{\pm{}}_{\epsilon}
\end{equation}

  Here, $A_{FB}$ is the forward-backward production asymmetry, and $A^{\pm{}}_{\epsilon}$ is the asymmetry due to different detection efficiencies for positively and negatively charged pions. Both can be eliminated through a relative measurement of $A_{CP}$ if the charged final-state particles are identical. The $CP$ asymmetry of the signal mode can then be expressed as: $A_{CP}(sig) = A_{\rm raw}(sig) - A_{\rm raw}(norm) + A_{CP}(norm)$. For $A_{CP}(norm)$, the world average value \cite{pdg} is used. Both the branching fraction and $A_{CP}$ are measured relative to other well measured decay channels.   Such  an  approach  enables  the  cancellation  of  several  sources  of  systematic  uncertainties that are common to both the signal and normalisation mode. All the analysis results presented here are based on  data collected by the Belle detector, operating at the asymmetric KEKB $e^+e^-$ collider ~\cite{abashian}.

\section{Search for radiative and hadronic decays $D^0 \rightarrow V\gamma$}
The radiative charm decay, $D^0 \rightarrow \phi\gamma$ has been first observed by the Belle Collaboration ~\cite{1}. In a subsequent analysis, the BaBar Collaboration measured both the decay, $D^0 \rightarrow \phi\gamma$ and $D^0 \rightarrow \bar{K^*}^0\gamma$ ~\cite{2}. The current world average values of the branching fractions are (2.70 $\pm$ 0.35) $\times 10^{-5}$ ($\phi$ mode) and (32.7 $\pm$ 3.4) $\times 10^{-5}$ ($\bar{K^*}^0$ mode) ~\cite{3}. The decay,  $D^0 \rightarrow \rho^0\gamma$ has not been observed up to date; the established upper limit by CLEO II is $\mathcal{B}$($D^0 \rightarrow \rho^0\gamma$) < 24 $\times 10^{-5}$ at 90\% confidence level ~\cite{4}. As radiative charm decays are dominated by non-perturbative long range dynamics, measurements of branching fractions can be a useful test for the QCD based theoretical calculations of the branching fractions. Further motivation for a study of radiative decays $D^0 \rightarrow V\gamma$, where $V$ is a vector meson, is that these decays could be sensitive to New Physics (NP) via CP asymmetry ($A_{CP}$ ). Theoretical studies ~\cite{5}~\cite{6} predict that in the SM extensions with chromomagnetic dipole operators, $A_{CP}$ can rise to several percent for $V$ = $\phi$ , $\rho^0$ , compared to the $\mathcal{O}(10 ^{-3})$ SM expectation. However, there has been no study of CP violation in $D^0 \rightarrow V\gamma$ decays conducted up to this point. The preliminary results of the measurement of the branching fractions and CP asymmetries in decays $D^0 \rightarrow V\gamma$ are presented here, where $V$ = $ \phi , \bar{K^*}^0 , \rho^0$ . This is the first observation of the decay $D^0 \rightarrow \rho^0\gamma$. The analysis is based on 943 fb$^{-1}$ of data collected by the Belle detector, operating at the asymmetric KEKB $e^+e^-$ collider ~\cite{7}.  The chosen normalisation modes are decays that feature the same charged final state particles as the signal decay. The  signal  decays  are  reconstructed  in  the  following  sub-decay  channels  of  the  vector  meson: $\phi \rightarrow K^+K^-$, $\bar{K^*}^0 \rightarrow K^-\pi^+$ and $\rho^0 \rightarrow \pi^+\pi^-$. In accordance, the corresponding normalisation modes are $D^0 \rightarrow K^+K^-$ ($\phi$ mode),  $D^0 \rightarrow K^-\pi^+$ ($\bar{K^*}^0$ mode) and $D^0 \rightarrow \pi^+\pi^-$ ($\rho^0$ mode).

\begin{figure}
\begin{center}
\includegraphics[width=4.6cm]{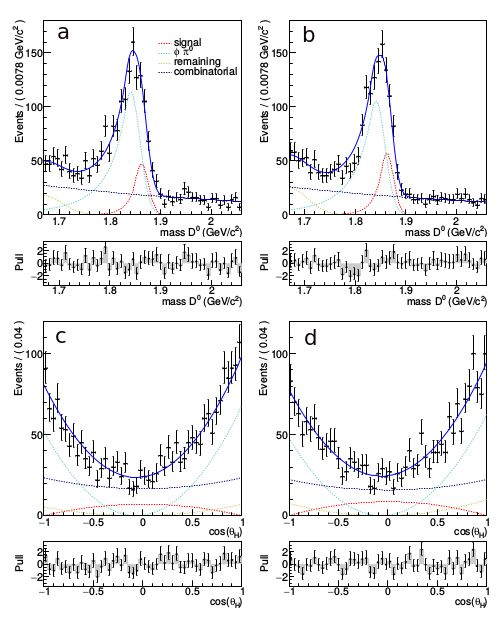} 
\includegraphics[width=4.6cm]{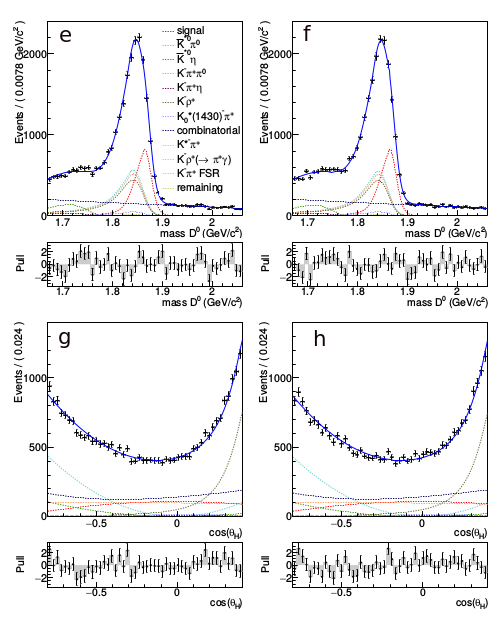} 
\includegraphics[width=4.6cm]{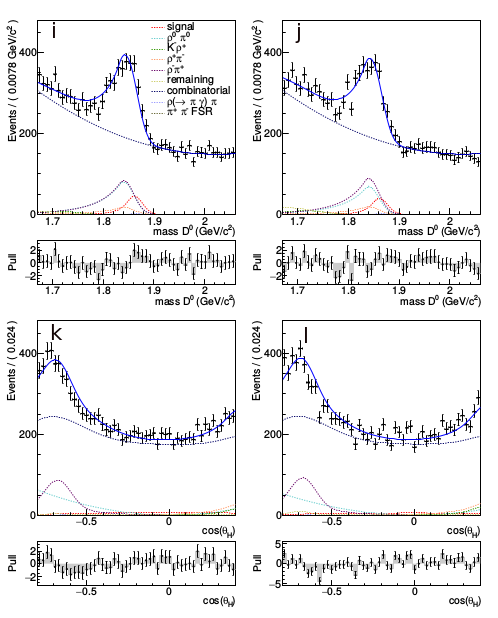} 
\caption{ The $M(D^0)$(top row) and cos$\theta_{H}$(bottom row) distributions for (a-d) the $\phi$ mode, (e-h) the $\bar{K^*}^0$ mode and (i-l) the $\rho^0$ mode,  in each $D^0$(left) and $\bar{D}^0$(right), with fit results superimposed.}
\label{fig:total}
\end{center}
\end{figure}
\subsection{Signal extraction and Systematic uncertainties}
Signal is extracted via a simultaneous 2-dimensional fit in the variables $m(D^0)$ and the cosine of the helicity angle (cos$\theta_{H}$), defined as the angle between the $D^0$ and positively (or negatively) charged hadron in the rest frame of the $V$ meson.  The background arises from decays $\pi^0$ to a pair of photons are suppressed ~\cite{tara}. The fit results are shown in Figure~\ref{fig:total} for the $\phi,\bar{K^*}^0$ and $\rho^0$ modes, respectively. The signal component is denoted with the dashed red line.  The extracted signal yields are 524$\pm$35 ($\phi$ mode), 9104$\pm$396 ($\bar{K^*}^0$ mode) and 500$\pm$85 ($\rho^0$ mode).  The extracted raw asymmetries are $-$0.091 $\pm$ 0.066 ($\phi$ mode), $-$0.002 $\pm$ 0.020 ($\bar{K^*}^0$ mode) and 0.064 $\pm$ 0.151 ($\rho^0$ mode). Here, the uncertainties are statistical only. The reconstruction efficiencies are 9.7\% ($\phi$ mode), 7.8\% ($\bar{K^*}^0$ mode) and 6.8\% ($\rho^0$ mode).
All the sources of systematic uncertainties are summarized in Ref.~\cite{tara}.
\subsection{Results}
The preliminary results are 
\\$\cal{B}$($D^{0} \rightarrow \phi\gamma$)  = (2.76 $\pm$ 0.20 $\pm$ 0.08) $\times 10^{-5}$,   $A_{CP}$($D^{0} \rightarrow \phi\gamma$)=  $-$0.094 $\pm$ 0.066$\pm$ 0.001,
\\$\cal{B}$($D^{0} \rightarrow \bar{K}^{*0}\gamma$) = (4.66  $\pm$ 0.21$\pm$ 0.18) $\times 10^{-4}$,  $A_{CP}$($D^{0} \rightarrow \bar{K}^{*0}\gamma$) = $-$0.003 $\pm$ 0.020$\pm$ 0.000,
\\$\cal{B}$($D^{0} \rightarrow \rho^{0}\gamma$) =  (1.77 $\pm$ 0.30 $\pm$ 0.08)$\times 10^{-5}$, $A_{CP}$($D^{0} \rightarrow \rho^{0}\gamma$) = $+$ 0.056 $\pm$ 0.151$\pm$ 0.006

where the first uncertainty is statistical and the second is systematic.  The branching fraction result of the $\phi$ mode is improved compared to the previous Belle result and is consistent with the world average value ~\cite{pdg}. The branching fraction of the $\bar{K}^{*0}$ mode reported here is 3.3$\sigma$ away from the result of the BaBar analysis. For the $\rho^{0}$ mode, this analysis reports the first observation of the decay. The significance of the observation is greater than 5$\sigma$, including systematic uncertainties. We also report the first-ever measurement of $A_{CP}$ in the decays $D^0 \rightarrow V\gamma$. Results are consistent with no CP asymmetry in any of the $D^0 \rightarrow V\gamma$ decay modes.
\section{Search for $D^0 \rightarrow K^0_S K_S^0$  decay}
\begin{figure}
\centering
\includegraphics[width=0.45\textwidth]{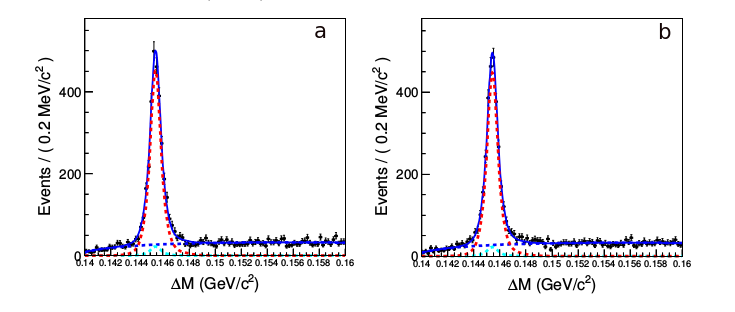} 
\includegraphics[width=0.42\textwidth]{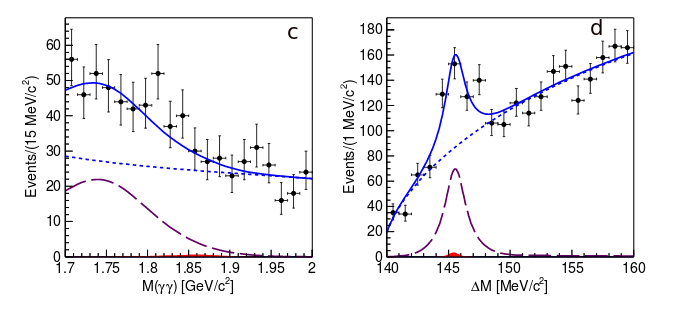}
\caption{(a,b), Distribution of the mass difference $\Delta M$ for the $K_S^0 K_S^0$, left(right) for the $D^{*+}$ ($D^{*-}$). Points with error bars are the data, the solid curves show the results of the fit, dashed (blue) curves are the background predictions and solid (cyan) is for the peaking background. (c,d), 2-dimensional fit in $M(D^0$)(left) and $\Delta M$(right). Blue (purple) dashed line denotes the combinatorial (peaking) background, while red histogram shows the signal component for the $D^0 \rightarrow \gamma\gamma$.}
\label{fig:markern}
\end{figure}
The $D^{0} \to K^0_{S} K^0_{S}$ decay is Single Cabibbo Suppressed channel ~\cite{hiller}. The most recent SM-based analysis obtained a 95\% confidence level upper limit of 1.1\% for direct $CP$ violation in this decay~\cite{Nierste:2015zra}. The search for $CP$ asymmetry in $D^{0} \to K^0_{S} K^0_{S}$ has been performed first by the CLEO Collaboration~\cite{Bonvicini:2000qm} using a data sample of 13.7 fb$^{-1}$  as $(-23 \pm 19)\%$. Recently LHCb measured time-integrated $CP$ asymmetry in $D^{0}\rightarrow K^0_{S}K^0_{S}$ as ($-$2.9 $\pm$ 5.2 $\pm$ 2.2)\%, where the first uncertainty is statistical and the second is systematic~\cite{Aaij:2015fua}. The LHCb result is consistent with no CPV, in agreement with SM expectations. We present here the preliminary result of the measurement of the  CP asymmetry in $D^0 \rightarrow K^0_{S}K^0_{S}$ decays using 921 fb$^{-1}$ data collected at the Belle detector. 

\subsection{Signal extraction and Systematic uncertainties}

Signal is extracted via a simultaneous fit of the $\Delta M $, where $\Delta M $ is the mass difference between reconstructed $D^*$ and $D^0$. Here the normalization mode is $D^0 \to K_S^0 \pi^0$. The $\Delta M $ distribution for the signal mode $D^0 \to K_S^0 K_S^0$ is shown in Figure~\ref{fig:markern}. 
The signal yield for $D^0 \to K_S^0 K_S^0$ is $5,399 \pm 87$ events and for $D^0 \to K_S^0 \pi^0$ is $531,807 \pm 796$ events. A simultaneous fit of the $\Delta M$ for $D^{*+}$ and $D^{*-}$ is used (Figure~\ref{fig:markern}) to estimate the asymmetry. 
The $A_{\rm raw}$ observed in data for $D^{0}\rightarrow K_{S}^{0}K_{S}^{0}$ and $D^{0}\rightarrow K_{S}^{0}\pi^{0}$ decay mode are $(+0.45 \pm 1.53)\%$ and $(+0.16 \pm 0.14)\%$, respectively.
We identify four sources of systematic uncertainty. The first is due to the uncertainties of the signal shapes, second we correct for a non-vanishing asymmetry originating from the different strong interaction of $K^0$ and ${\bar{K}}^0$ mesons with nucleons of the detector material, estimated to be $-0.11\%$ 
and assign an additional systematic uncertainty of $0.01\%$. Third, the peaking background yield is determined and fixed from the $K_S^0$ mass sideband. Fourth, the fit procedure is repeated with its yield varied by its statistical uncertainty. Similarly, for its raw asymmetry, fixed from the $K_S^0 \pi^0$ measurement. 
The total systematic uncertainty for the time integrated CP violating asymmetry $A_{CP}$ in the $D^{0}\rightarrow K_{S}^{0}K_{S}^{0}$ decay is $\pm$0.17\% \cite{nibedita}. 

\subsection{Results}
The measured time-integrated $CP$-violating asymmetry $A_{CP}$ in the $D^{0}\rightarrow K_{S}^{0}K_{S}^{0}$ decay is found to be
$A_{CP}$ = ($-$0.02 $\pm$ 1.53 $\pm$ 0.17) \% using a data sample of 921~fb$^{-1}$ integrated luminosity. The dominant systematic uncertainty comes from the $A_{CP}$ error of the normalization channel. The result is consistent with SM expectations and is a significant improvement compared to the previous measurements of CLEO~\cite{Bonvicini:2000qm} and LHCb Collaborations~\cite{Aaij:2015fua}, already probing the region of interest.

\section{Search for rare decay $D^0 \rightarrow \gamma\gamma$}
We also present a search for the rare charm decay $D^0 \rightarrow \gamma\gamma$ ~\cite{nisar}. This decay has not been observed yet. The best upper limit to date was set by BaBar at $\mathcal{B}$($D^0 \rightarrow \gamma\gamma$) < 2.2 $\times 10^{-6}$ (90\% C.L.) ~\cite{9}. The decay $D^0 \rightarrow \gamma\gamma$ represents a good probe for NP as the SM prediction for the branching fraction, which is of the order of $\mathcal{O}$(10$^{-8}$), can be enhanced by several orders of magnitude by NP contributions. The Minimal Supersymmetric Standard Model suggests that the exchange of gluinos can enhance the $D^0 \rightarrow \gamma\gamma$ branching ratio up to 6 $\times 10 ^{-6}$ ~\cite{10}~\cite{11}. The present measurement, conducted on 832 fb$^{-1}$ of Belle data, represents the most stringent upper limit for this decay. Peaking  background  arises  from  decays  comprising  a $\pi^0$ and/or $\eta$ meson, which decays to a pair of photons: $D^0 \rightarrow \pi^0\pi^0$, $D^0 \rightarrow \eta\pi^0$,$D^0 \rightarrow \eta\eta$, $D^0 \rightarrow K_S^0(\pi^0\pi^0)\pi^0$, $D^0 \rightarrow K_L^0\pi^0$. These background types are suppressed with a $\pi^0(\eta)$ veto and suppression of merged ECL clusters through the ratio $\frac{E_9}{E_{25}}$. The branching fraction is calculated relative to the normalisation mode $D^0 \rightarrow K_S^0\pi^0$.

\subsection{Signal extraction and Systematic uncertainties}
Signal is extracted through a 2-dimensional fit of m($D^0$) and $\Delta m$ as shown in the Figure~\ref{fig:markern}. The efficiency is 7.3\% and we extract a signal yield of 4 $\pm$ 15 events. 
The efficiency is 7.2\% and we extract a yield of 343050 $\pm$ 673 events for the normalisation channel.
The systematic uncertainties are summarized in Ref.~\cite{nisar}. 
\subsection{Results}
In the absence of signal, a frequentist method is used to estimate the upper limit on the branching fraction at a 90\% confidence level. The final result after including systematic uncertainties is $\mathcal{B}(D^0 \rightarrow \gamma\gamma) < 8.4 \times 10^{-7}$ at 90\% C.L. \cite{nisar}. This result represents the world's most stringent upper limit to date.


\section{Acknowledgments}
We thank the KEKB group, all institutes and agencies that have support the work of the members of the Belle Collaboration. 



\begin{thebibliography}{99}
\bibitem{61} The charge conjugation is applied throughout unless stated explicitly.
\end{thebibliography}
\end{document}